\documentclass[letterpaper,english,reprint, aps]{revtex4-1}
\usepackage[T1]{fontenc}
\usepackage[latin9]{inputenc}
\usepackage{babel}
\usepackage{amsmath}
\usepackage{amssymb}
\usepackage{graphicx}
\usepackage[unicode=true]
 {hyperref}

\makeatletter

\providecommand{\tabularnewline}{\\}

\makeatother

\begin{document}
\title{Walker diffusion method for solution of ohmic circuit problems}
\author{Clinton DeW. Van Siclen}
\email{cvansiclen@gmail.com}

\address{1435 W 8750 N, Tetonia, Idaho 83452, USA}
\date{June 24, 2019}
\begin{abstract}
A probabilistic method is derived for solution of ohmic circuit problems.
It is compared to the standard approach, which is construction and
solution of a set of coupled, linear equations manifesting Kirchhoff's
laws. An example is made of an electrical circuit that has the complicated
connectivity of a bond-and-node Sierpinski triangle, which would be
tedious to solve by matrix methods.
\end{abstract}
\maketitle
The walker diffusion current density $\mathbf{J}(\mathbf{r})$ at
location $\mathbf{r}$ due to the driving force $-\nabla\rho(\mathbf{r})$
(the walker density gradient) is
\begin{equation}
\mathbf{J}(\mathbf{r})=-D(\mathbf{r})\,\nabla\rho(\mathbf{r})
\end{equation}
where $D(\mathbf{r})$ is the local walker diffusion coefficient.
This rendering of Fick's first law for particle diffusion resembles
Ohm's law for the electrical current density $J_{i\rightarrow j}$
from node $i$ to node $j$ due to the electrical potential difference
$\phi_{j}-\phi_{i}$:
\begin{equation}
J_{i\rightarrow j}=-\sigma_{ij}\left(\phi_{j}-\phi_{i}\right)/r_{ij}
\end{equation}
where $\sigma_{ij}$ is the conductivity of the bond connecting the
nodes and $r_{ij}$ is the bond length. Clearly the local walker diffusion
coefficient $D(\mathbf{r})$ and the local walker density $\rho(\mathbf{r})$
in Fick's law correspond to the conductance $\sigma_{ij}/r_{ij}\equiv g_{ij}$
and the electrical potential $\phi_{i}$, respectively, in Ohm's law.
This motivates the development of a walker diffusion method for solution
of the \textit{set} of Ohm's law equations that represents an electrical
circuit.

The key criterion on walker behavior is that for a system at \textit{equilibrium}
(that is, no walker flux) all walker densities $\left\{ \rho_{i}\right\} $
must be constant and identical. Thus the probability $p_{i\rightarrow j}$
that a walker at node $i$ moves to the connected node $j$ on its
next move attempt must satisfy the relation $p_{i\rightarrow j}=p_{j\rightarrow i}$,
which implies $p_{i\rightarrow j}\propto g_{ij}$.

The equilibrium condition is achieved by the ``variable residence
time'' algorithm obtained as follows: On average, an attempted move
by the walker at node $i$ is successful with probability $\pi_{i}=\sum p_{i\rightarrow k}\propto\sum g_{ik}$
where the sums are taken over all connected nodes $k$. {[}Note that
the walker is a ``blind ant'': the probability of success increases
with the number of connected nodes $k$.{]} Then the time interval
associated with a \textit{successful} move is, on average, $T_{i}=\tau/\pi_{i}$
where $\tau$ is the time interval associated with an \textit{attempted}
move. That successful move is made to connected node $j$ (rather
than to a different connected node) with probability $P_{i\rightarrow j}=p_{i\rightarrow j}/\sum p_{i\rightarrow k}=g_{ij}/\sum g_{ik}$.
Thus the actual behavior of the walker is well approximated by a sequence
of moves in which the destination of each move from a node $i$ is
determined randomly by the set of probabilities $\left\{ P_{i\rightarrow j}\right\} $,
where
\begin{equation}
P_{i\rightarrow j}=\frac{g_{ij}}{\sum_{k}g_{ik}}
\end{equation}
and the time interval over which the move occurs is
\begin{equation}
T_{i}=\frac{\tau}{\pi_{i}}\propto\frac{1}{\sum_{k}g_{ik}}.
\end{equation}
The sums in these expressions are taken over all nodes $k$ that are
connected to node $i$. As the walker moves over the system of nodes,
the time interval $T_{i}$ associated with a visit to node $i$ is
accrued to the ``residence time'' $t_{i}$. Then the (normalized)
walker density $\rho_{i}$ at node $i$ is
\begin{equation}
\rho_{i}=\frac{t_{i}}{\left\langle t_{k}\right\rangle }
\end{equation}
where the average value $\left\langle t_{k}\right\rangle $ is taken
over all nodes $k$ comprising the circuit.

The application of a potential difference $\triangle V$ between two
nodes of the circuit (causing an electrical current to flow from one
to the other) corresponds to the selection of one node (designated
by the subscript $\alpha$) to be a walker \textit{source} and the
other (designated by the subscript $\beta$) to be a walker \textit{sink}.
A large number of walkers, placed at the source node, diffuse over
the circuit in the manner described above until they visit the sink
node (where no residence time is accrued so that $\rho_{\beta}=0$).
The residence times at the nodes then give the \textit{steady-state}
distribution $\left\{ \rho_{i}\right\} $, from which the electrical
potentials are obtained according to
\begin{equation}
\phi_{i}=\left(\rho_{i}/\rho_{\alpha}\right)\triangle V=\left(t_{i}/t_{\alpha}\right)\triangle V
\end{equation}
where $\rho_{\alpha}$ is the walker density at the source node.

The conventional method to obtain the potentials (and currents $I$)
in an ohmic electrical circuit is by solving a set of linear equations,
each of which corresponds to Kirchhoff's current law for one of the
nodes of the circuit. By conservation of charge, the sum of currents
\textit{to} and \textit{from} a node must be zero. Thus in the case
of node $i$,
\begin{equation}
\sum_{k}I_{ik}=0
\end{equation}
where $I_{ik}$ is the current through the bond connecting nodes $i$
and $k$, and the sum is over all nodes $k$ connected to the node
$i$. Equation (7) for node $i$ is equivalently
\begin{equation}
\sum_{k}g_{ik}\left(\phi_{i}-\phi_{k}\right)=0
\end{equation}
which simplifies to
\begin{equation}
\phi_{i}-\sum_{k}P_{i\rightarrow k}\,\phi_{k}=0
\end{equation}
where the sum is over all nodes $k$ connected to node $i$. Note
however that equations for the \textit{source} and \textit{sink} nodes
are, instead, $\phi_{\alpha}=\triangle V$ and $\phi_{\beta}=0$.
The set of equations (of number equal to the number $n$ of nodes
in the circuit) is then solved for the set of potentials $\left\{ \phi_{i}\right\} $,
from which the currents are obtained.

Solution is accomplished most conveniently by putting the relations
in matrix form: $\mathbf{A}\mathbf{x}=\mathbf{b}$ where $\mathbf{x}$
is a column vector comprised of the (unknown) potentials $\phi_{i}$
and $\mathbf{A}$ is an $n\times n$ matrix whose entries are constructed
from the set $\left\{ g_{ij}\right\} $ of conductances. The tedious
task is filling out the matrix $\mathbf{A}$ to reproduce the set
of equations, but this approach can use standard matrix solvers and
gives ``exact'' values for the potentials.

In contrast, the walker diffusion method (WDM) is easily implemented
in a very simple computer code, and requires as input only a description
of the circuit. That is most conveniently given by an $n\times n$
array where the entry $(i,j)$ is the conductance $g_{ij}$. (Of course
when two nodes are not connected, the corresponding entry is $0$.)
As this is a probabilistic method, the set $\left\{ \phi_{i}\right\} $
of electrical potentials \textit{approaches} the true values as the
number of walkers released at the source node increases.

An example of a complicated bond-and-node circuit is shown in Fig.
(1). Whatever the distribution of values $\left\{ g_{ij}\right\} $,
a single walker (or multiple walkers) diffusing over the closed system
produces $\rho_{i}\rightarrow1$ at all nodes.
\begin{figure}[t]
\includegraphics[scale=0.5]{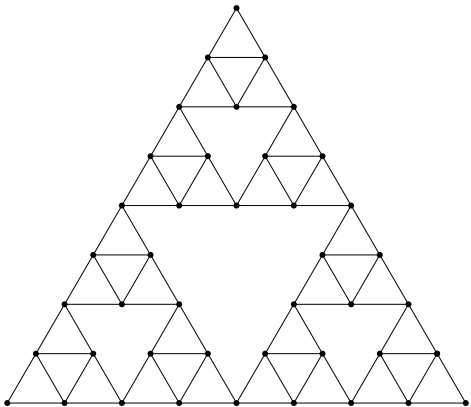}

\caption{Resistor network having the connectivity of a Sierpinski triangle.
The equivalent resistance for a potential drop across two apex vertices
is calculated by the WDM.}

\end{figure}

Note that this circuit has the form of a Sierpinski triangle (AKA
gasket or sieve) at iteration $3$. When a potential difference $\triangle V$
is applied across two of the three corner nodes (apex vertices), an
analytical value for the equivalent resistance $R_{\alpha\beta}$
can be obtained by use of the triangle-star ($\triangle\rightarrow Y$)
transformation \citep{key-1}. In the case that all bonds in the Sierpinski
triangle have conductance $g_{ij}=1$, the resistance $R_{\alpha\beta}=(2/3)(5/3)^{3}\approx3.08642$.

The WDM obtains $R_{\alpha\beta}$ in the following way: The walker
flux emitted at the source node is $J_{\alpha}^{(w)}=\sum_{k}g{}_{\alpha k}\left(\rho_{\alpha}-\rho_{k}\right)$
which corresponds to the electrical current $I_{\alpha}=\sum_{k}g_{\alpha k}\left(\phi_{\alpha}-\phi_{k}\right)$,
where the sum (in both expressions) is over all nodes $k$ connected
to the source node $\alpha$. Then the equivalent resistance $R_{\alpha\beta}=\triangle V/I_{\alpha}=\rho_{\alpha}/J_{\alpha}^{(w)}$.

Table I gives calculated values of $R_{\alpha\beta}$ for the particular
case mentioned above, showing that the true value is approached as
the number $N$ of walkers released at the source node increases.

\begin{table}[h]
\caption{Calculated resistance $R_{\alpha\beta}$ after $N$ walks between
source node and sink node. Note that a different sequence of values
would occur for a different initial random ``seed''. The true value
$R_{\alpha\beta}\approx3.08642$.}
\begin{tabular}{ccccccc}
\hline 
 & $N$ &  &  &  & $R_{\alpha\beta}$ & \tabularnewline
\hline 
\hline 
 & $10^{5}$ &  &  &  & $3.05995$ & \tabularnewline
\hline 
 & $10^{6}$ &  &  &  & $3.07489$ & \tabularnewline
\hline 
 & $10^{7}$ &  &  &  & $3.08236$ & \tabularnewline
\hline 
 & $10^{8}$ &  &  &  & $3.08610$ & \tabularnewline
\hline 
\end{tabular}

\end{table}

Note that another way to obtain the set $\left\{ \phi_{i}\right\} $
(and so the electrical currents $\left\{ I_{i\rightarrow j}\right\} $
as well) is via the relation
\begin{equation}
J_{i\rightarrow j}^{(w)}=g_{ij}\left(\rho_{i}-\rho_{j}\right),
\end{equation}
where $J_{i\rightarrow j}^{(w)}$ is the walker flux between connected
nodes $i$ and $j$. A large number of walks between source node $\alpha$
and sink node $\beta$ produce the fluxes $\left\{ J_{i\rightarrow j}^{(w)}\right\} $.
Then by use of Eq. (10) the walker densities $\left\{ \rho_{i}\right\} $
are calculated one at a time beginning with $\rho_{\beta}=0$ and
working up to $\rho_{\alpha}$. Finally the $\left\{ \rho_{i}\right\} $
are converted to $\left\{ \phi_{i}\right\} $ by multiplying the former
by $\triangle V/\rho_{\alpha}$ according to Eq. (6).

\section*{Concluding remarks}

The walker diffusion method utilizes the ``variable residence time''
algorithm for walker diffusion over a bond-and-node undirected network,
where the node degree may vary dramatically over the network (so the
network is not necessarily a regular grid). {[}The \textit{degree}
of node $i$ is the number of nodes $k$ to which it is connected.{]}

It should be evident that the WDM is \textit{not} a particle model
of a physical process: a randomly diffusing walker does not resemble
an electron responding to an applied electric field. Rather, the eponymous
walker diffuses over the nodal circuit according to particular rules,
thereby ``solving'' the system of local Ohm's law equations associated
with the set of nodes.

Despite the attractive physical arrangement of the nodes in Fig. (1),
it is only their \textit{connectivity} that defines the circuit. Thus
there is no length scale, or Euclidean dimension within which the
circuit is embedded. As a consequence there can be no diffusion coefficient
$D_{w}$ that describes the behavior of the walker. This is in contrast
to the \textit{object} that is a bond-and-node Sierpinski triangle
embedded in 2D space, considered in Ref. \citep{key-2}.
\begin{acknowledgments}
I thank Professor Robert ``Bob'' Smith (Department of Geological
Sciences) for arranging my access to the resources of the University
of Idaho Library (Moscow, Idaho).
\end{acknowledgments}

\end{document}